\documentclass{article} 
\usepackage{iclr2027_conference,times}

\usepackage{amsmath,amsfonts,bm}

\def\eqref#1{equation~\ref{#1}}

\def\1{\bm{1}}

\DeclareMathAlphabet{\mathsfit}{\encodingdefault}{\sfdefault}{m}{sl}
\SetMathAlphabet{\mathsfit}{bold}{\encodingdefault}{\sfdefault}{bx}{n}

\iclrfinalcopy
\usepackage{hyperref}
\usepackage{url}
\usepackage[utf8]{inputenc} 
\usepackage[T1]{fontenc}    
\usepackage{hyperref}       
\usepackage{url}            
\usepackage{booktabs}       
\usepackage{amsfonts}       
\usepackage{nicefrac}       
\usepackage{microtype}      
\usepackage{xcolor}         
\usepackage{subcaption}
\usepackage{graphicx}
\title{Learning Perceptual Rules \\ in a Fragmented Visual World}

\author{Tianqin Li, Isaac Lin, Leiran Song, Tai Sing Lee\\\\
Computer Science Department\\
Center for the Neural Basis of Cognition\\ 
Carnegie Mellon University\\
Pittsburgh, PA 15213, USA \\\\
\texttt{\{tianqinl, isaacl, leirans, taislee\}@cs.cmu.edu}
}

\begin{document}

\maketitle

\begin{abstract}
Visual perception organizes incomplete sensory evidence into coherent structures, but where do the principles governing this organization come from? We investigate whether learning to predict missing visual information can give rise to structural priors resembling classical Gestalt principles. Masked autoencoders (MAEs) trained on natural images exhibit perceptual completion consistent with good continuation, symmetry, and global form. Critically, when local evidence surrounding a missing region is held constant, manipulating distant elements systematically changes its reconstruction, demonstrating that local predictions are constrained by the global organization of the visual input. We next manipulate visual experience by adapting MAEs to either natural images or texforms that preserve local image statistics while disrupting global spatial organization. This altered experience weakens globally supported completion and shifts perceptual inference toward local continuation when local and global constraints conflict. Together, these results suggest that perceptual organization reflects structural regularities of natural scenes acquired through visual prediction: predictions integrate structural priors across multiple spatial scales, and the balance between local and global organization adapts to the statistical structure of visual experience. More broadly, our findings suggest that classical Gestalt principles may emerge as adaptive perceptual priors learned from the statistical structure of the visual world.
\end{abstract}

\section{Introduction}

\label{sec:introduction}

A missing part of an image can often be filled in several ways.
Nearby edges may suggest continuing a line, while the arrangement
of the whole object may suggest a different shape. For example,
a gap in one side of a cross can be filled by extending a straight
boundary or restoring an arm that matches the opposite side.
Both completions fit the visible pixels, but express different
expectations about how the parts belong together. Gestalt
principles such as good continuation, symmetry, and closure
describe regularities in this organization. We ask whether
learning to reconstruct natural images supports such structural
expectations, and how local and global relationships constrain
predictions about missing content.

Masked autoencoders (MAEs) provide a computational setting for
addressing this question. They encode visible image patches
and reconstruct the missing regions, creating an opportunity
to learn predictive relationships at multiple spatial scales.
Nearby fragments can support contour continuation, while distant
parts can constrain a larger configuration. The reconstructed
pixels make these expectations observable: we can examine
which missing structures the model generates and which visible
relationships influence them. We refer to learned tendencies
favoring particular spatial relationships as \emph{structural
priors}, without assuming explicit probabilistic representations
or symbolic Gestalt rules.

An organized reconstruction alone does not reveal whether it
depends on local interpolation or broader context. We therefore
probe MAEs with controlled contour, symmetry, and Kanizsa-type
stimuli, then manipulate distant elements while preserving the
visible neighborhood of the missing region. In symmetric dot
displays, moving a distant counterpart shifts the reconstructed
element, while disrupting the surrounding symmetry weakens the
response. In contour displays, rotating distant segments changes
the target reconstruction despite unchanged neighboring fragments.
These interventions show that the tested completions depend on
spatial relationships beyond the local evidence held constant
across conditions.

We next ask how the model resolves competition between local
and global constraints. Following the logic of local--global
completion conflicts \cite{carrigan2016differentiating}, we use
cross and lobed-circle stimuli for which local continuation and
global symmetry predict different missing structures. The
alternative completions are compatible with identical visible
pixels, allowing us to measure which interpretation the
reconstruction more closely follows. This comparison tests
the content of structural predictions, beyond whether a
completed boundary is strong or weak.

To test the role of global structure in the training data,
we continually pretrain an MAE on texforms that disrupt object
organization while retaining selected texture and coarse-form
information \cite{wen2023does}. Continued training from the same
checkpoint on matched natural images serves as a control.
Texform adaptation weakens boundary completion and reduces
geometric agreement with the tested Kanizsa contours.
It also shifts reconstruction toward local continuation in
both cue-conflict shape families, reversing the average global
preference for cross stimuli. These results suggest that
coherent training-image structure supports globally organized
completion. Because the intervention starts from a pretrained
model, it tests changes to existing structural expectations
rather than their initial emergence.

Finally, decoder-head interventions connect these behaviors
to internal computation: ablating heads identified through
activation replacement weakens contour completion, while
individual-head rankings vary across tasks and evaluation
metrics. Together, the experiments establish two complementary
findings: distant context changes predictions under fixed local
evidence, and disrupting training-image structure changes
predictions under identical test inputs. They support an account
of Gestalt-like completion in which learned regularities about
relationships among image parts constrain the reconstruction
of missing information.

\section{Related Work}
\label{sec:related_work}

\paragraph{Perceptual organization and completion.}
Human perceptual completion depends on local contour geometry
\citep{kellman1991theory} and broader shape constraints
\citep{vanlier1995competing,liu1999role,carrigan2016differentiating}.
Studies of neural networks likewise reveal Gestalt-like responses,
but with substantial variation across architectures, stimuli,
and evaluation measures
\citep{baker2018deep,kim2021neural,biscione2023mixed,
zhang2026finding,li2025local}.
Video-prediction and predictive-coding networks also exhibit
illusory-contour responses, including effects in reconstructed
images \citep{lotter2020neural,pang2021predictive}.
Using MAE reconstruction \citep{he2022masked}, we examine how
distant structure influences missing-region predictions under
fixed local evidence, including conflicts between local
continuation and global symmetry.

\paragraph{Natural image structural prior.}
Natural-image statistics link contour regularities to perceptual
grouping \citep{geisler2001edge,elder2002ecological}.
Training-image transformations can also alter network cue
preferences, although texture robustness and global structural
sensitivity are distinct \citep{geirhos2019imagenet,wen2023does}.
Texforms disrupt recognizable object structure while retaining
selected texture and coarse-form information \citep{long2018mid}.
We use texform continued pretraining as a structural ablation,
with matched natural-image training as a control, to test how
disrupting global training structure modifies existing
completion behavior.

\paragraph{Mechanisms of completion.}
Component-level interventions have linked transformer attention
heads to contour continuity and object binding
\citep{tartaglini2026walk}, and to convex-completion preferences
in BEiT \citep{arnold2026locating}.
We extend this analysis to MAE decoder computation through
activation replacement, ablation on held-out contour stimuli,
and comparisons of individual-head ablation effects across
completion tasks.

\section{Results}
\label{sec:results}

We investigate the structural relationships expressed in MAE
completions through three complementary approaches. First, we
characterize completion behavior using controlled artificial stimuli
and manipulate the visible context surrounding a missing region.
Second, we examine how these behaviors change following continued
pretraining on natural images versus texform images. Third, we
intervene on decoder attention heads to identify components that
contribute to contour completion, global symmetry, as well as Kanizsa square illusory boundaries and investigate whether these components specialize or overlap across different completion phenomena.

\subsection{Structural preferences in masked completion}
\label{sec:completion_behavior}

Masked reconstruction provides a direct means of examining the
structural expectations expressed by a visual model. When image
content is hidden, the model must use visible evidence and learned
regularities to predict the missing region. We first examine whether
these predictions exhibit recognizable completion preferences,
including continuation between aligned fragments and boundary
completion in Kanizsa-type configurations.

\begin{figure}[ht]
    \centering
    \includegraphics[width=0.9\linewidth]{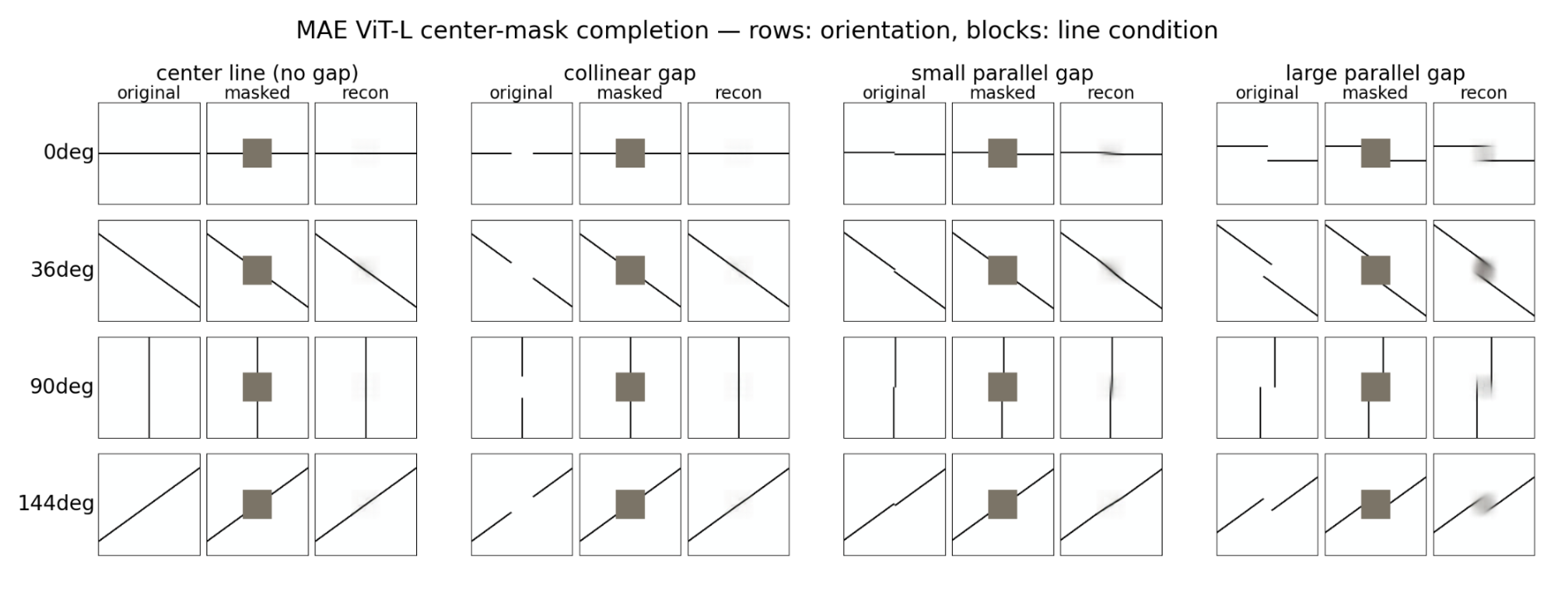}
    \vspace{-3mm}
    \caption{
        Contour completion under variations in orientation and
        alignment. MAE reconstructs a centrally masked region
        for four stimulus conditions: a continuous line, a collinear
        gap, and small and large perpendicular offsets between
        parallel segments. Rows show stimulus orientations; each
        block presents the original image, masked input, and
        reconstruction. Gray squares indicate the masked region.
        Collinear configurations generally yield more coherent
        line-like continuations, whereas larger offsets often
        produce diffuse or displaced marks.
    }
    \label{fig:mae_line_continuation_comp1}
    \vspace{-3mm}
\end{figure}

\paragraph{Contour continuation.}
We present MAE with line stimuli containing a centrally masked
region and vary the orientation and relative alignment of the visible
segments. The conditions include a continuous line, a collinear gap,
and small or large perpendicular offsets between parallel segments
(Figure~\ref{fig:mae_line_continuation_comp1}).

Collinear configurations generally produce more coherent line-like
reconstructions through the masked region. Larger offsets often
produce diffuse, displaced, or less continuous predictions.
Reconstruction contrast and sharpness also vary with orientation.
These observations indicate that the geometry of the visible
fragments influences completion and are consistent with a preference
for good continuation. The following experiments examine whether
completion also depends on relationships beyond the immediately
neighboring fragments.

\vspace{-3mm}




\subsection{Symmetric context guides the completion of missing elements}
\label{sec:symmetry_completion}

We next ask whether a missing element is reconstructed according to
its relationship with a spatially separated counterpart. We use
dot arrangements with vertical reflection symmetry and mask one
element while leaving its counterpart visible. Within each layout,
the mask and its visible local neighborhood remain identical across
conditions.

\begin{figure}[ht]
    \centering
    \vspace{-3mm}
    \includegraphics[width=1\linewidth]{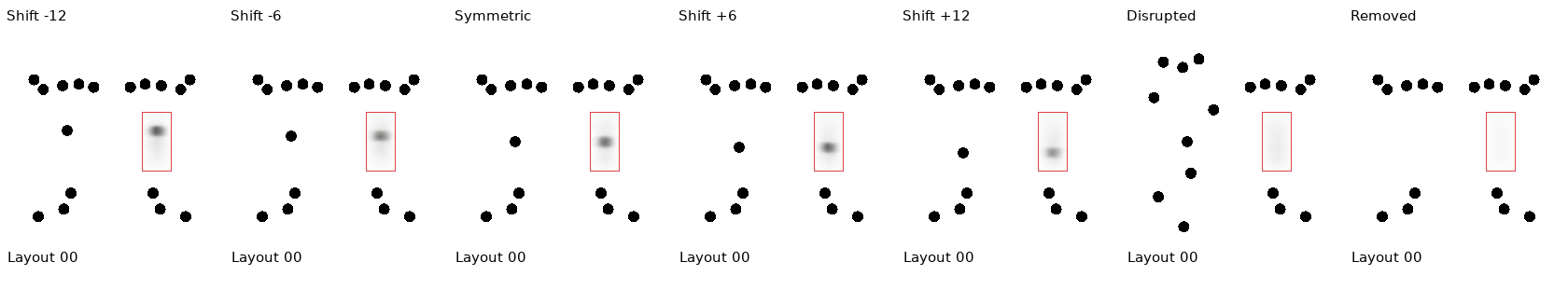}
    \caption{\textbf{Completion of a missing element in symmetric dot arrangements.}
  Rows show an example layout with vertical reflection symmetry.
  Columns show counterpart displacements of $-12$, $-6$, $0$, $+6$, and $+12$ pixels along the symmetry axis, followed by disrupted-context and counterpart-removed
  controls.
  Each panel displays the predicted pixels within the masked region.
  Dots have a diameter of 12 pixels.
  The mask and its visible local neighborhood remain identical across conditions within each layout.
  The reconstructed response follows the displaced counterpart and becomes weaker when the surrounding symmetry is disrupted or the counterpart is removed.
  All panels use a common $[0,1]$ intensity scale without per-panel contrast normalization.}
    \label{fig:placeholder}
\end{figure}

We displace the visible counterpart along the symmetry axis by
$-12$, $-6$, $0$, $+6$, or $+12$ pixels. We additionally compare
conditions in which the surrounding symmetric arrangement is
disrupted or the counterpart is removed.

In the illustrated layouts, the reconstructed response shifts with
the displaced counterpart. Its strength decreases when the
surrounding symmetry is disrupted or the counterpart is removed.
These observations indicate that the prediction depends on both
the distant counterpart and the configuration in which it appears.
They motivate a distinction between simple transfer of an isolated
element and completion supported by a broader symmetric arrangement.

\subsection{Distant contour organization modulates local completion}
\label{sec:context_completion}

The preceding experiment concerns relationships between
corresponding elements. We next examine whether the organization
of a surrounding contour influences reconstruction when the
visible evidence adjacent to the target remains unchanged.

\paragraph{Distant orientation perturbations.}
We construct segmented contour displays and mask a target segment.
The two segments neighboring the target are held fixed, while five
distant segments are rotated about their centers. The perturbations
range from the aligned configuration to rotations of
$\pm15^\circ$, $\pm30^\circ$, $\pm45^\circ$, and $90^\circ$.

\begin{figure}[ht]
    \centering
    \begin{subfigure}{0.9\linewidth}
        \centering
        \includegraphics[width=\linewidth]{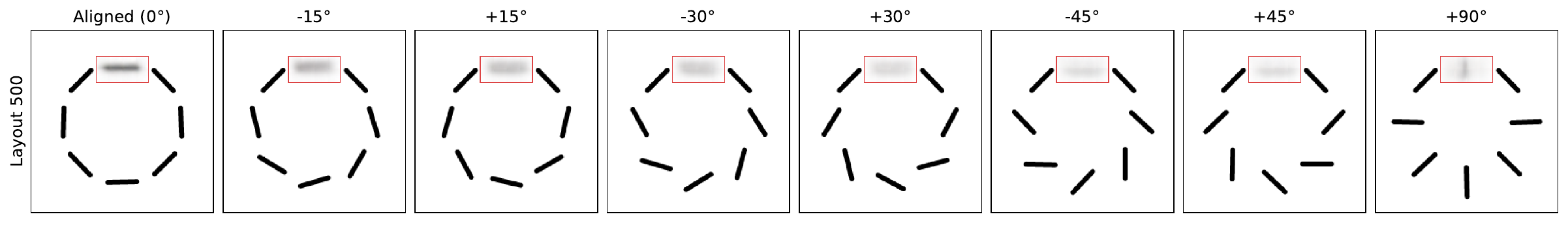}
        \label{fig:top}
    \end{subfigure}


    \begin{subfigure}{0.9\linewidth}
        \centering
        \includegraphics[width=\linewidth]{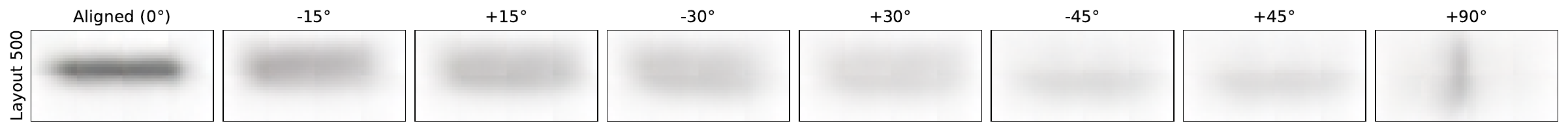}
        \label{fig:bottom}
    \end{subfigure}
    \vspace{-3mm}
    \caption{\textbf{Reconstruction of a missing segment under distant orientation perturbations.}
  Top: stimulus context with the predicted target region inserted; red boxes mark the masked region.
  Bottom: enlarged reconstructions of the same target region.
  Columns indicate the rotation applied to five distant segments about their fixed centers.
  The two segments neighboring the target remain unchanged, preserving the visible local context.
  The demo uses 6-pixel-wide segments.}
    \label{fig:combined}
\end{figure} 

Despite identical neighboring segments, the reconstructed target
changes in contrast and spatial structure as distant segment
orientations change. In the illustrated layouts, the aligned
configuration produces a more coherent contour-like prediction,
whereas several perturbed configurations yield weaker or more
diffuse responses. Thus, the neighboring fragments alone do not
determine the model's completion.

\paragraph{Distant contour fragmentation.}
We further vary the fragmentation of an irregular surrounding
contour while preserving a fixed masked target gap, its endpoints,
their tangent directions, and the visible local neighborhood.
Conditions contain $0$, $1$, $2$, $4$, $6$, $8$, or $12$ additional
distant openings.

\begin{figure}[t]
    \centering
    \begin{subfigure}{0.8\linewidth}
        \centering
        \includegraphics[width=\linewidth]{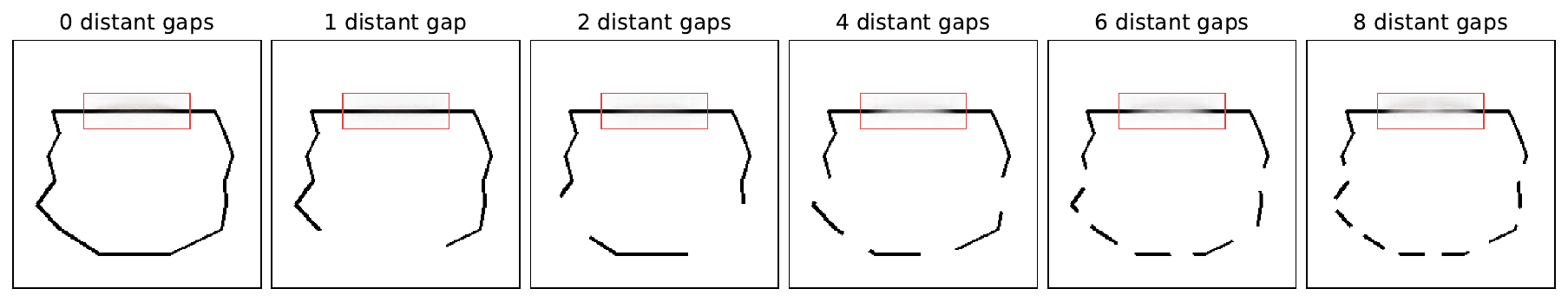}
        \label{fig:top}
    \end{subfigure}
    \begin{subfigure}{0.8\linewidth}
        \centering
        \includegraphics[width=\linewidth]{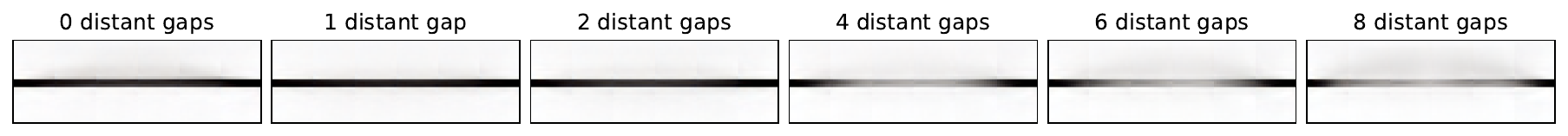}
        \label{fig:bottom}
    \end{subfigure}
    \vspace{-3mm}
    \caption{\textbf{Completion of an irregular contour with increasing fragmentation.}
  Columns show 0, 1, 2, 4, 6, or 8 distant openings, in addition to a fixed 96-pixel masked target gap.
  The target endpoints, tangent directions, and visible local neighborhood are identical across conditions.
  Panels show an illustrative irregular layout and its reconstructed target region. 
  All reconstructions use a common $[0,1]$ display scale without per-panel contrast normalization.}
    \label{fig:combined}
\end{figure}

The model continues to reconstruct the target contour across these
conditions, although the appearance and strength of the prediction
vary. The displayed examples do not establish a simple monotonic
relationship between gap count and completion strength. Instead,
they show that local continuation can persist despite substantial
fragmentation of the surrounding contour. This manipulation probes
the contribution of enclosing context, but does not by itself
isolate a preference for closure from changes in contour support
and gap geometry.

Together, these experiments show that completion is sensitive to
some changes in distant organization while remaining robust to
others. They provide a basis for characterizing when broader
context contributes beyond the visible fragments nearest the
missing region.

\subsection{Disrupting training-image structure weakens structural completion}
\label{sec:texform_training}

\begin{figure}[htbp]
    \centering
    \begin{subfigure}[c]{0.45\linewidth}
        \centering
        \includegraphics[width=\linewidth]
            {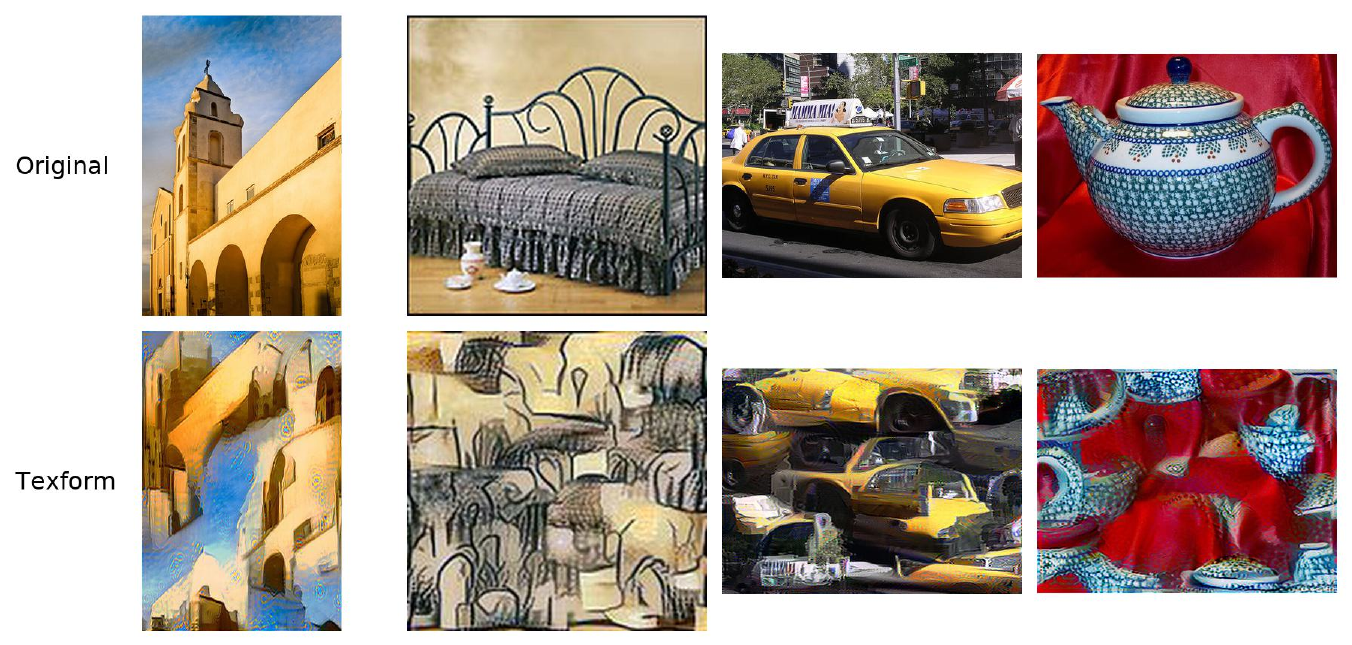}
    \end{subfigure}%
    \hspace{0.015\linewidth}%
    \begin{subfigure}[c]{0.5\linewidth}
        \centering
        \includegraphics[width=\linewidth]
            {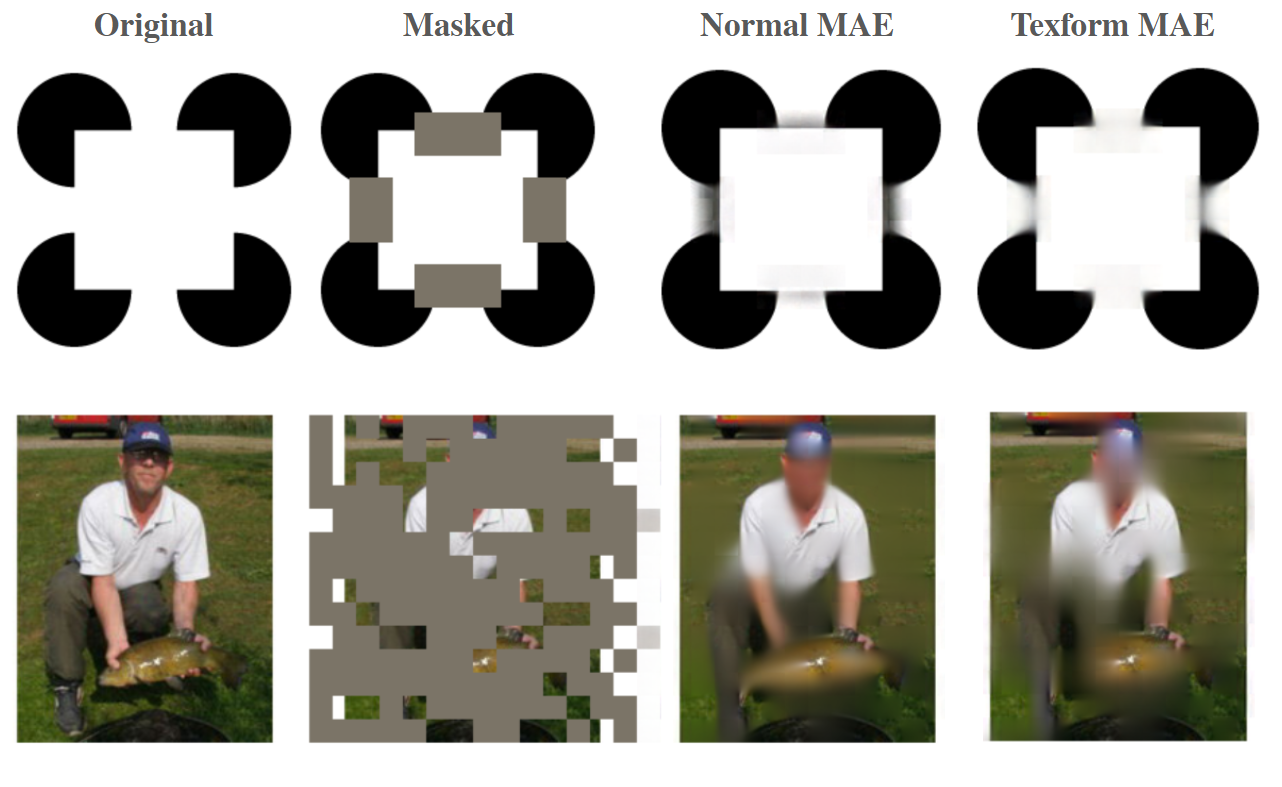}
            
    \end{subfigure}

    \caption{Left: Matched natural images (top) and texforms
    (bottom), generated following \cite{wen2023does}.
    Separate MAE-B models underwent 500 epochs of continued
    pretraining on these image sets.
    Right (Top): Kanizsa-square reconstructions, showing weaker
    boundary-like completion after texform adaptation than
    after natural-image adaptation in the illustrated examples. Right (Bottom): In the natural-image example, the texform-adapted model
produces a less coherent reconstruction of the person's arm,
with a less distinct connection to the torso than the
naturally adapted model. This is consistent with weaker
completion of structure supported by the surrounding body
configuration.}
    \label{fig:texform_training}
\end{figure}

We next ask whether structural completion depends on the organization
of the images encountered during continued training. We use texform
images to disrupt global object structure while preserving selected
local image statistics, including aspects of texture and color,
as well as some coarse-form information
(Figure~\ref{fig:texform_training}, left).
Starting from the same pretrained MAE-B checkpoint, we continually
pretrain separate models on matched sets of 100,000 natural images
or their texform counterparts for 500 epochs. At evaluation, both
models receive identical visible inputs and masks. We also evaluate
the original MAE-B without continued pretraining as a reference.

Figure~\ref{fig:texform_training} (right) illustrates differences
in completion following the two training conditions.
For the Kanizsa square, the texform-adapted model produces weaker
boundary-like structure between the inducers than the naturally
adapted model. A related difference appears in the natural-image
example: natural-image adaptation yields a more recognizable arm
extending from the torso, whereas texform adaptation produces
a less coherent arm reconstruction. These examples suggest that
disrupting structure in the training images weakens the reconstruction
of missing parts consistent with the surrounding configuration,
across both artificial and natural stimuli. They motivate the
following systematic evaluation of structural completion.

\subsection{Visual experience shapes the completion of convex and concave boundaries}
\label{sec:convex_concave_completion}

We next examine whether the effects of visual experience extend
to the geometry of completed boundaries. A missing contour can
bend outward or inward relative to a central region, producing
a convex or concave shape. We therefore extend the straight-sided
Kanizsa square to configurations whose visible inducer boundaries
support convex or concave continuations. This manipulation asks
whether reconstruction follows the geometry suggested by the
visible fragments, and whether this correspondence changes after
continued pretraining on natural images or texforms.

\begin{figure}[ht]
\vspace{-3mm}
    \centering
    \includegraphics[width=\linewidth]{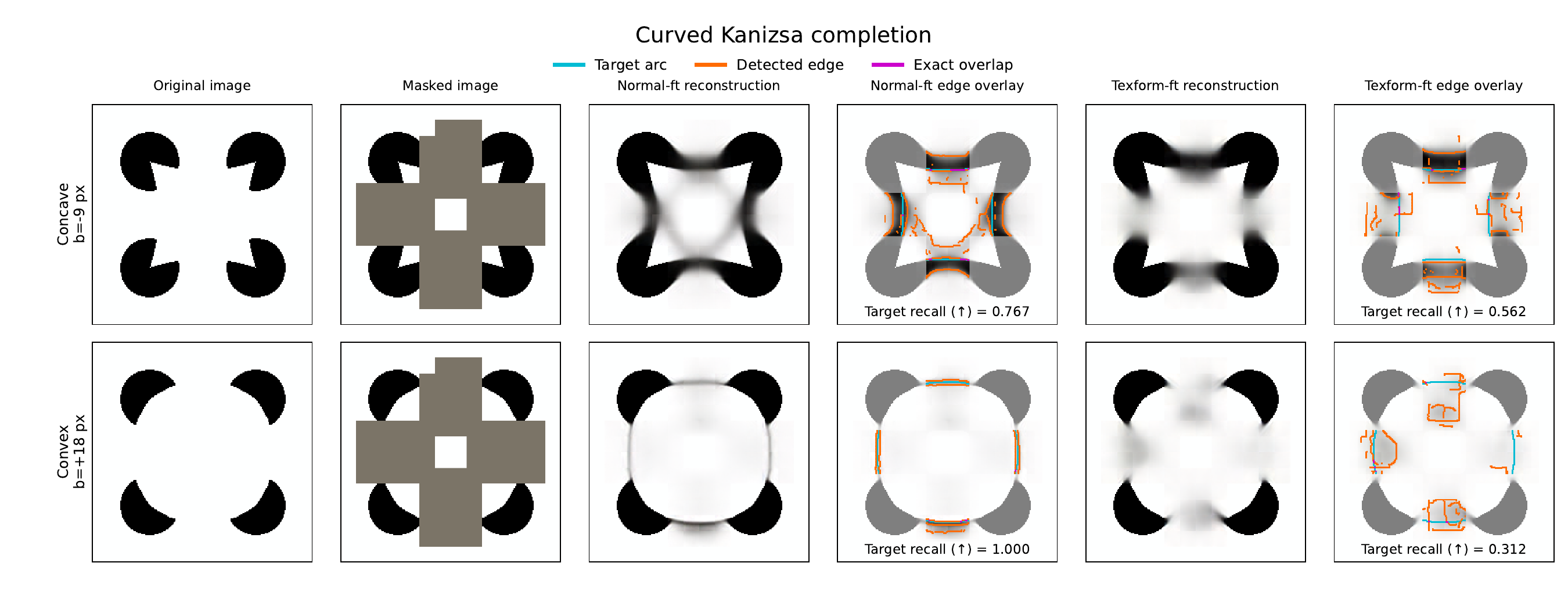}
    \caption{Completion of concave, straight, and convex boundaries
    after natural-image versus texform adaptation.
    Inducer radius is fixed at $r=30$ pixels; rows vary boundary
    bulge $b\in\{-9, 18\}$ pixels (More systemtic variation can be found in Appendix Figure~\ref{fig:convex_concave_ksquare_complete}).
    Columns show the original stimulus, masked input,
    natural-adapted reconstruction and edge overlay, and
    texform-adapted reconstruction and edge overlay.
    Cyan marks target contours, orange detected edges, and
    magenta exact pixel overlap. Target recall uses a two-pixel
    matching tolerance. Natural-image adaptation yields higher
    target recall in the illustrated curved configurations;
    both models attain full recall for the straight boundary.}
    \label{fig:convex_concave_ksquare}
    \vspace{-3mm}
\end{figure}

We evaluate 20 configurations combining four inducer radii
$r\in\{24,30,36,42\}$ pixels with five boundary patterns, using
identical inputs and masks across models. The inter-inducer gap
is fixed at 48 pixels, and boundary bulge is defined as
$b=\operatorname{round}(\alpha s)$, where $s=48+2r$ and
$\alpha\in\{-1/6,-1/12,0,1/12,1/6\}$; negative, zero, and positive
values specify concave, straight, and convex boundaries,
respectively. Figure~\ref{fig:convex_concave_ksquare} shows
representative concave and convex configurations at $r=30$
pixels, with the full sweep in
Appendix Figure~\ref{fig:convex_concave_ksquare_complete}.
Natural-image adaptation produces boundaries that broadly
follow the target geometry, whereas texform adaptation yields
more fragmented or displaced edges, particularly for the
strongly convex configuration. Both models nevertheless attain
full target recall for the straight-boundary example in the
appendix, suggesting that texform adaptation impairs recovery
of curved geometry without eliminating boundary completion.

\paragraph{Quantifying contour geometry and strength.}
We evaluate reconstructed boundaries against generator-defined
target contours within the four masked inter-inducer gaps,
excluding mask seams and using identical evaluation regions
across models and boundary patterns at each radius.
We measure boundary F1 with a two-pixel tolerance, symmetric
nearest-edge distance capped at 16 pixels, and boundary strength
as the absolute normal image gradient within two pixels of
the target contour (For more details, please refer to Appendix~\ref{app:boundary-metrics}).

\begin{table}[ht]
  \centering
  \caption{Edge reconstruction quality on curved Kanizsa stimuli.
  Values are mean $\pm$ sample standard deviation across 20 stimuli
  (four inducer radii $\times$ five boundary patterns, including the
  straight-edge condition), without head ablation.
  Bold indicates the best mean in each column.}
  \label{tab:kanizsa_edge_quality}
  \begin{tabular}{lccc}
  \toprule
  Model & Boundary F1 $\uparrow$
        & Boundary distance (px) $\downarrow$
        & Edge strength $\uparrow$ \\
  \midrule
  Normal-ft
      & $\mathbf{0.516} \pm 0.150$
      & $\mathbf{4.086} \pm 1.860$
      & $\mathbf{0.152} \pm 0.077$ \\
  Texform-ft
      & $0.263 \pm 0.129$
      & $6.293 \pm 1.065$
      & $0.043 \pm 0.027$ \\
  \bottomrule
  
  \end{tabular}
  \vspace{-2mm}
  \end{table}

Averaged across 20 configurations
(mean $\pm$ sample SD), natural-image adaptation yields higher
F1 ($0.516 \pm 0.150$ versus $0.263 \pm 0.129$),
lower distance ($4.086 \pm 1.860$ versus $6.293 \pm 1.065$ pixels),
and greater strength ($0.152 \pm 0.077$ versus $0.043 \pm 0.027$)
than texform adaptation. These results indicate experience-dependent
changes in contour completion systemtically.
Variability reflects stimulus differences, not independent
training runs.

\subsection{Competing local and global cues reveal experience-dependent completion preferences}
\label{sec:global_local_conflict}

\begin{figure}[ht]
    \centering
    \includegraphics[width=0.9\linewidth]{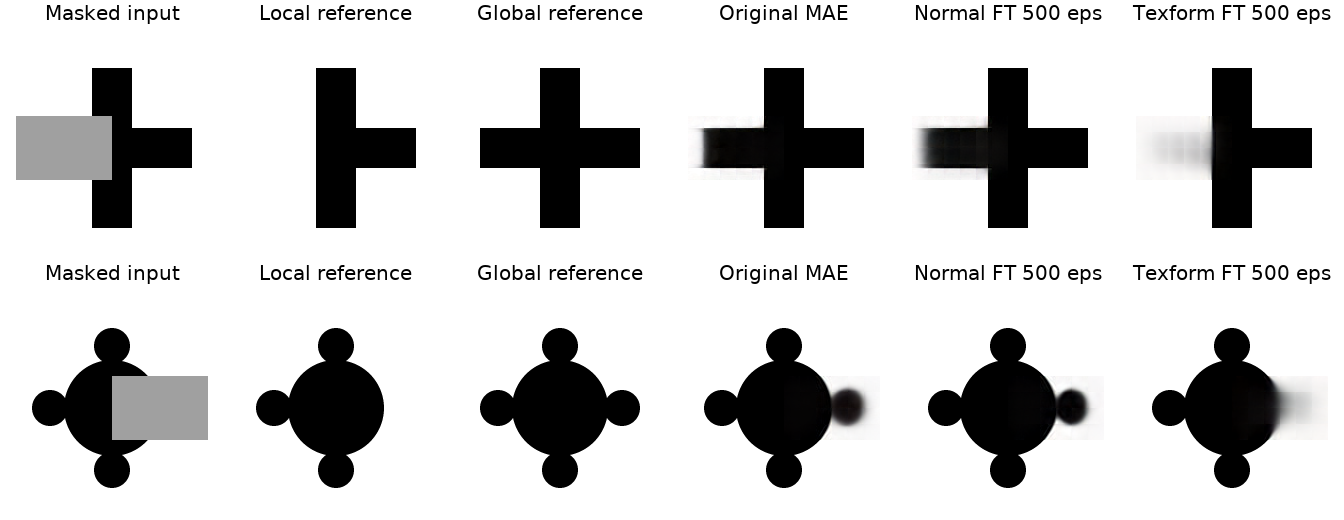}
    \caption{Local–global conflicts in MAE reconstruction. The top row shows the cross stimulus, and the bottom row shows the lobed stimulus rotated by 180°.
  Columns show the masked input, the local-continuation reference, the global-symmetry reference, and reconstructions from the original pretrained MAE-B, the MAE-B
  continually pretrained on natural images for 500 epochs (“Normal FT 500 eps”), and the MAE-B continually pretrained on texform images for 500 epochs (“Texform FT 500
  eps”). Gray regions indicate patches withheld from the encoder. All models receive the same visible image and mask; reconstruction panels combine decoder
  predictions inside the mask with the unchanged visible pixels. Stimuli are schematic adaptations of Carrigan, Palmer, and Kellman (2016, Figure 5).}
    \label{fig:global_local}
\end{figure}

Having examined completion under individual structural cues,
we next ask how visual experience shapes reconstruction when
local continuation and global symmetry favor different outcomes.
We use cross and lobed-circle stimuli adapted from
\cite{carrigan2016differentiating}
(Figure~\ref{fig:global_local}). Local continuation predicts a
straight boundary in the cross and a smooth circular boundary
in the lobed shape, whereas global symmetry predicts an
additional arm or lobe matching its visible counterpart.
The two reference completions differ only within the mask,
making both compatible with identical visible evidence.

We compare the original MAE with models continually pretrained
for 500 epochs on natural images or matched texforms.
Each receives identical inputs and masks across 72 variants
per shape (three sizes, three mask widths, four rotations,
and two contrast polarities). We measure masked-region RGB
MSE against each reference; lower error indicates closer
agreement with that completion.
Table~\ref{tab:cue_conflict_errors} reports means and sample
standard deviations across stimulus variants.


\begin{table}[htbp]
  \centering
  \caption{Masked-region MSE relative to local and global
  completions (lower is better). Values are mean $\pm$ sample SD
  across 72 variants per shape; SD reflects stimulus variation,
  not training-run uncertainty. Bold marks the lower mean per row.
  Adapted models received 500 epochs of continued pretraining.}
  \label{tab:cue_conflict_errors}
  \vspace{-3mm}
  \begingroup
  \setlength{\tabcolsep}{7pt}
  \renewcommand{\arraystretch}{1.1}
  \scalebox{0.9}{%
    \begin{tabular}{llcc}
      \toprule
      Shape & Model & Local MSE $\downarrow$ & Global MSE $\downarrow$ \\
      \midrule
      Cross
        & Original MAE
        & $0.2245 \pm 0.0968$ & $0.0496 \pm 0.0475$ \\
        & Natural-adapted
        & $0.2690 \pm 0.0947$ & $\mathbf{0.0731} \pm 0.0469$ \\
        & Texform-adapted
        & $\mathbf{0.0074} \pm 0.0075$ & $0.2468 \pm 0.0853$ \\
      \midrule
      Lobed circle
        & Original MAE
        & $0.0735 \pm 0.0531$ & $\mathbf{0.0504} \pm 0.0554$ \\
        & Natural-adapted
        & $0.0617 \pm 0.0490$ & $0.0666 \pm 0.0634$ \\
        & Texform-adapted
        & $\mathbf{0.0218} \pm 0.0166$ & $0.1224 \pm 0.0703$ \\
      \bottomrule
    \end{tabular}%
  }
  \endgroup
  \vspace{-3mm}
\end{table}

For the cross, both the original and naturally adapted models
favor the global reference, whereas texform adaptation reverses
this preference: local-reference MSE falls to 0.0074, compared
with global-reference MSE of 0.2468. The lobed shape shows a
weaker initial global preference, near-equal reference errors
after natural-image adaptation, and a clearer local preference
after texform adaptation (local MSE 0.0218 versus global MSE
0.1224). Thus, baseline preferences depend on shape, but
texform adaptation shifts both families toward the local
reference, with lower local-reference error and higher
global-reference error than either comparison model.
These findings extend the preceding boundary-completion results:
training experience changes not only the strength of reconstructed
structure, but also its relative agreement with competing
local and global interpretations.

\subsection{Decoder attention heads contribute to contour completion}
\label{sec:decoder_interventions}
\begin{figure}[ht]
    \centering
    \vspace{-3mm}
    \includegraphics[width=0.8\linewidth]{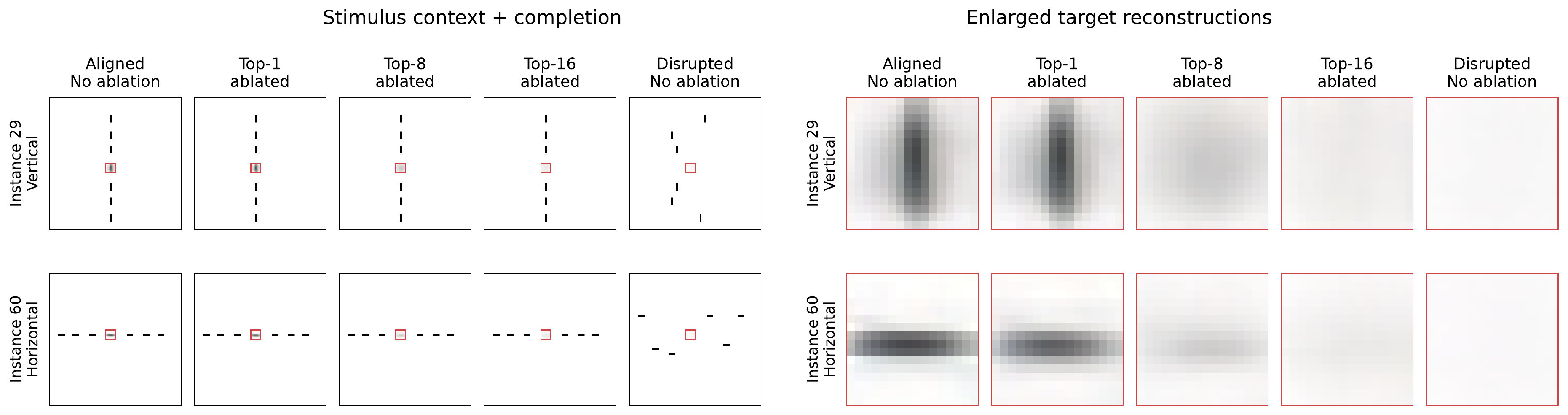}
    \caption{\textbf{Effect of ranked decoder-head ablation on contour completion.}
      Rows show horizontal and vertical examples (held-out instances 60 and 29) for MAE-B (finetuned with normal nature images for 500 epochs).
      Columns show the aligned stimulus without ablation, the aligned stimulus with the top 1, 8, or 16 decoder heads ablated, and the disrupted stimulus without ablation. Heads were ranked separately using bidirectional activation replacement on a separate discovery set.
      Ablation sets selected head outputs to zero at all decoder token positions before the attention output projection.
      Increasing ablation weakens the reconstructed contour.
      Context panels retain the original stimulus pixels outside the central target; they are not full-image reconstructions.
      Examples were selected for visual clarity, and all predictions use the same $[0,1]$ display scale without per-panel contrast normalization.
     }
    \label{fig:ranked-decoder-ablation}
    \vspace{-3mm}
\end{figure}

We investigate decoder contributions to contour completion in
MAE-B (continual pretrained by normal images) by ranking its 128 attention heads
using bidirectional activation replacement on 24 discovery
instances. Scores combine the reduction in target reconstruction
error when aligned-condition activations replace disrupted ones
and the increase under the reverse replacement. With this ranking
fixed, we ablate the top $k$ heads
($k\in\{1,4,8,16,24,32\}$) by zeroing their outputs at all decoder
positions, including the class token, before the attention output
projection. Figure~\ref{fig:ranked-decoder-ablation} illustrates
the effects of ablating the top 1, 8, or 16 heads on two held-out
examples, while Table~\ref{tab:normal-ft-ranked-head-ablation}
reports results across all 40 held-out instances. Increasing
ablation weakens the illustrated contours and progressively
raises mean target reconstruction error, from
$55.96\times10^{-3}$ without ablation to $80.68\times10^{-3}$
after removing the top 32 heads, approaching the disrupted-input
reference of $81.69\times10^{-3}$.
Further details appear in
Appendix~\ref{sec:app_more_contour_completion_ablation}.

\begin{table}[htbp]
\centering\small
\setlength{\tabcolsep}{8pt}
\renewcommand{\arraystretch}{1.18}
\caption{
\textbf{Effect of ranked decoder-head ablation on contour completion.}
In normally fine-tuned MAE-B (epoch 500), the top $k$ decoder heads are zeroed during reconstruction of aligned contour stimuli. Head rankings are fixed using activation replacement on 24 discovery instances, with ablation evaluated on 40 held-out instances. Values report target-patch RGB MSE (mean $\pm$ bootstrap SE; $\times 10^{-3}$; lower is better), computed from raw, unclipped predictions. Standard errors use 2,000 paired bootstrap resamples. $\Delta$ measures the change from the unablated aligned baseline, with positive values indicating increased reconstruction error after ablation. The unablated displaced-input condition provides a separate reference for the effect of disrupting contour alignment.
}
\begin{tabular}{lrr}
\toprule
Input / ablated group & MSE & $\Delta$ vs. aligned \\
\midrule
Aligned, no ablation & $55.96 \pm 3.94$ & $0.00 \pm 0.00$ \\
Disrupted, no ablation & $81.69 \pm 5.38$ & \textemdash \\
\midrule
Aligned, top-1 & $60.37 \pm 3.91$ & $4.41 \pm 1.07$ \\
Aligned, top-4 & $70.15 \pm 4.28$ & $14.20 \pm 1.66$ \\
Aligned, top-8 & $74.41 \pm 4.52$ & $18.45 \pm 2.44$ \\
Aligned, top-16 & $77.34 \pm 4.95$ & $21.38 \pm 3.01$ \\
Aligned, top-24 & $79.02 \pm 4.37$ & $23.07 \pm 2.69$ \\
Aligned, top-32 & $80.68 \pm 4.33$ & $24.72 \pm 2.71$ \\
\bottomrule
\end{tabular}
\label{tab:normal-ft-ranked-head-ablation}
\end{table}

Two held-out examples---vertical instance 29 and horizontal
instance 60---show little visible change after single-head
ablation, but weaker contours and higher target error after
ablating 8 or 16 heads, relative to the intact aligned condition.
Context views and enlarged target patches illustrate these
effects alongside unablated aligned and disrupted references.
These examples link context-dependent completion to selected
decoder components.

\subsection{Decoder-head contributions vary across completion tasks}
\label{sec:decoder-head-specialization}

\begin{figure*}[ht]
    \centering
    \includegraphics[width=0.6\linewidth]{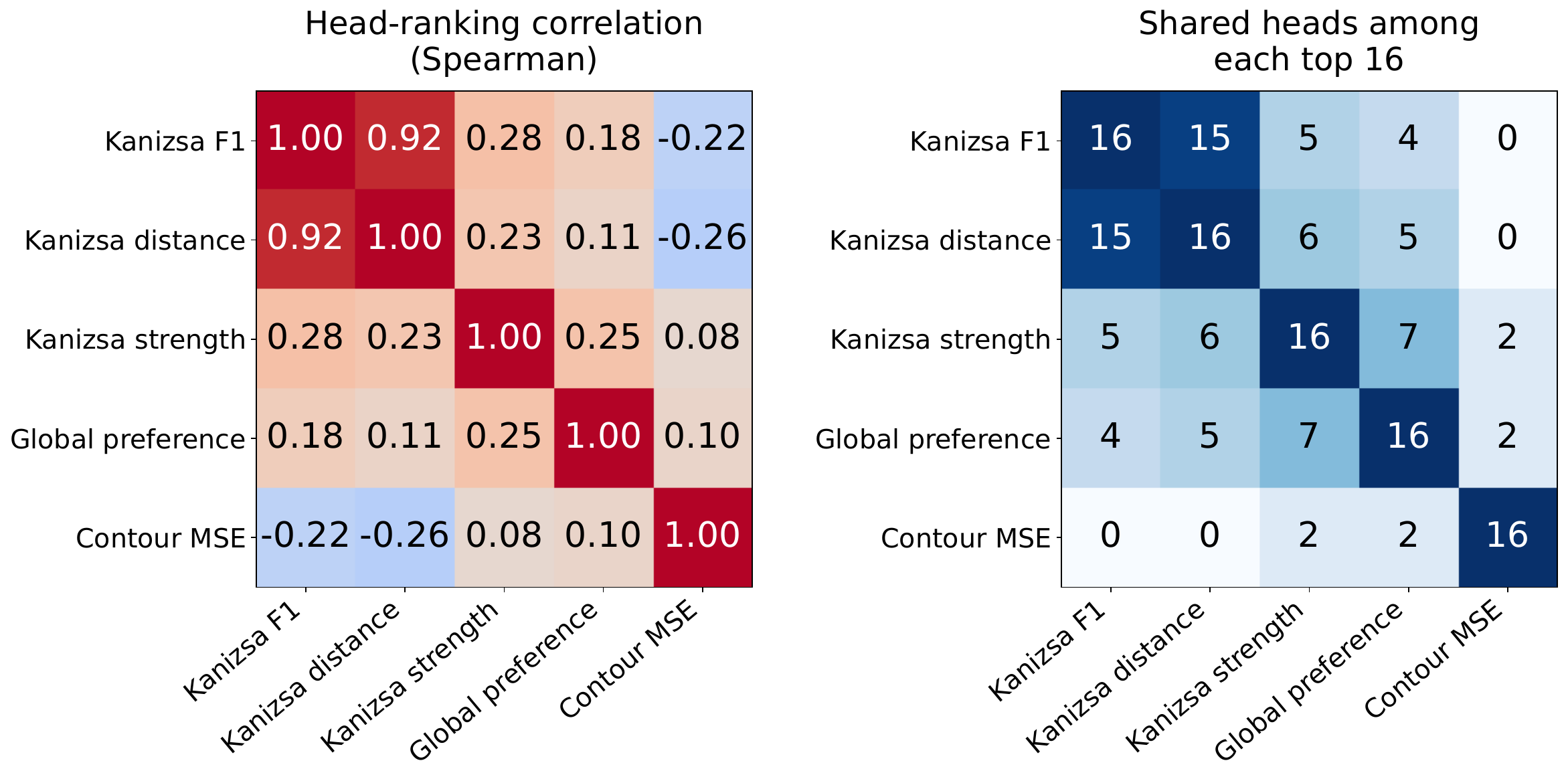}
    \caption{Agreement between decoder-head contribution rankings.
    Left: Spearman correlations across all 128 heads, ranked separately
    by mean individual-head ablation effect for each outcome.
    Right: numbers of shared heads between the corresponding top-16
    sets. Kanizsa F1 and boundary distance select similar heads,
    whereas agreement with contour MSE is substantially weaker.
    The three Kanizsa metrics measure different aspects of the same
    task. Both panels are descriptive comparisons; the overlap counts
    are not significance tests.}
    \label{fig:decoder-head-rank-agreement-main}
    \vspace{-3mm}
\end{figure*}

Do the same decoder heads support different forms of completion?
We individually ablate all 128 decoder heads in the naturally
adapted MAE-B and measure changes in curved Kanizsa completion,
local/global cue preference, and contour continuation.
Figure~\ref{fig:decoder-head-rank-agreement-main} compares head
rankings across these outcomes. Rankings agree strongly between
Kanizsa boundary F1 and distance (Spearman $\rho=0.924$;
15 shared top-16 heads), but differ across tasks:
contour-error rankings share no top-16 heads with either
Kanizsa boundary metric and only two with global preference.
Even within Kanizsa completion, edge-strength rankings agree
only modestly with boundary-accuracy rankings.
These results indicate task- and metric-dependent ablation
effects, although differences in stimuli and measurements
preclude interpreting these dissociations as dedicated
Gestalt-rule encoding. Full procedures and results appear in
Appendix~\ref{app:decoder-head-specialization}.
Grouping-related processing may also involve the encoder:
qualitative attention maps show interactions consistent with
contour grouping (Appendix~\ref{app:encoder_attention}),
suggesting that relevant spatial relationships may be
represented before decoding.

\section{Conclusion}
\label{sec:conclusion}
We examined how the spatial relationships learned by MAEs
are expressed in the reconstruction of missing image content.
Controlled stimuli reveal completion preferences consistent
with symmetry, good continuation, and Kanizsa configurations.
Changing distant structure while preserving the visible
neighborhood of the target changes these predictions,
showing that completion depends on relationships beyond
local image evidence.

Ablating global structure in the training images through
texform continued pretraining weakens structural completion
and shifts agreement toward local continuation under
competing local and global cues, relative to matched
natural-image training. Decoder-head interventions further
connect completion to internal computation, with ablation
effects varying across tasks and measured outcomes.

Together, these findings suggest that the global organization
of training images helps sustain learned constraints on how
visible parts predict missing structure. Reconstruction makes
these constraints observable, while interventions on context,
training-image structure, and decoder components reveal their
dependencies. This provides a computational account of
Gestalt-like completion as the expression of learned spatial
regularities, without requiring explicitly specified
perceptual rules.

\section*{Generative AI Disclosure}

We used generative AI tools (ChatGPT Astra) in this work
to assist with refining the conceptual framing and hypotheses,
providing feedback on experimental methodology, discussing the
interpretation of results, and helping implement analysis and
experimental code. We also used generative AI tools to help
identify relevant literature, improve the clarity and
readability of the manuscript, revise figure captions, and
assist with \LaTeX{} formatting.

We did not use generative AI tools for tasks such as
translation, qualitative or thematic data analysis, and other disclosure categories not relevant to this
work are not applicable.

All AI-assisted work was reviewed by the authors. In
particular, AI-assisted code was manually inspected and its
outputs were checked against the underlying experimental setup
and results. Literature suggestions and text edits were also
reviewed and revised as needed by the authors. We take full
responsibility for the final content of this work, including
all text, claims, code, and artifacts produced with the aid of
generative AI.

\bibliography{iclr2027_conference}
\bibliographystyle{iclr2027_conference}
\newpage
\newpage
\section{Appendix}

\subsection{Reproduciblity}
We utilize huggingface transformer library as well as the official MAE repo to download and analysis the internal representation of the MAE Models. All experiment is done on a single L40S GPU which has 48GB GRAM memory and 1T CPU RAM. The code will be hosted at \url{https://anonymous.4open.science/r/mae_learning_completion_rules-CAB4} with documentation of how to access the finetuned natural image and the generated texform images.

\subsection{Contour completion across orientations and alignments}
\label{app:line_continuation}

\begin{figure}[ht]
    \centering
    \includegraphics[width=0.9\linewidth]{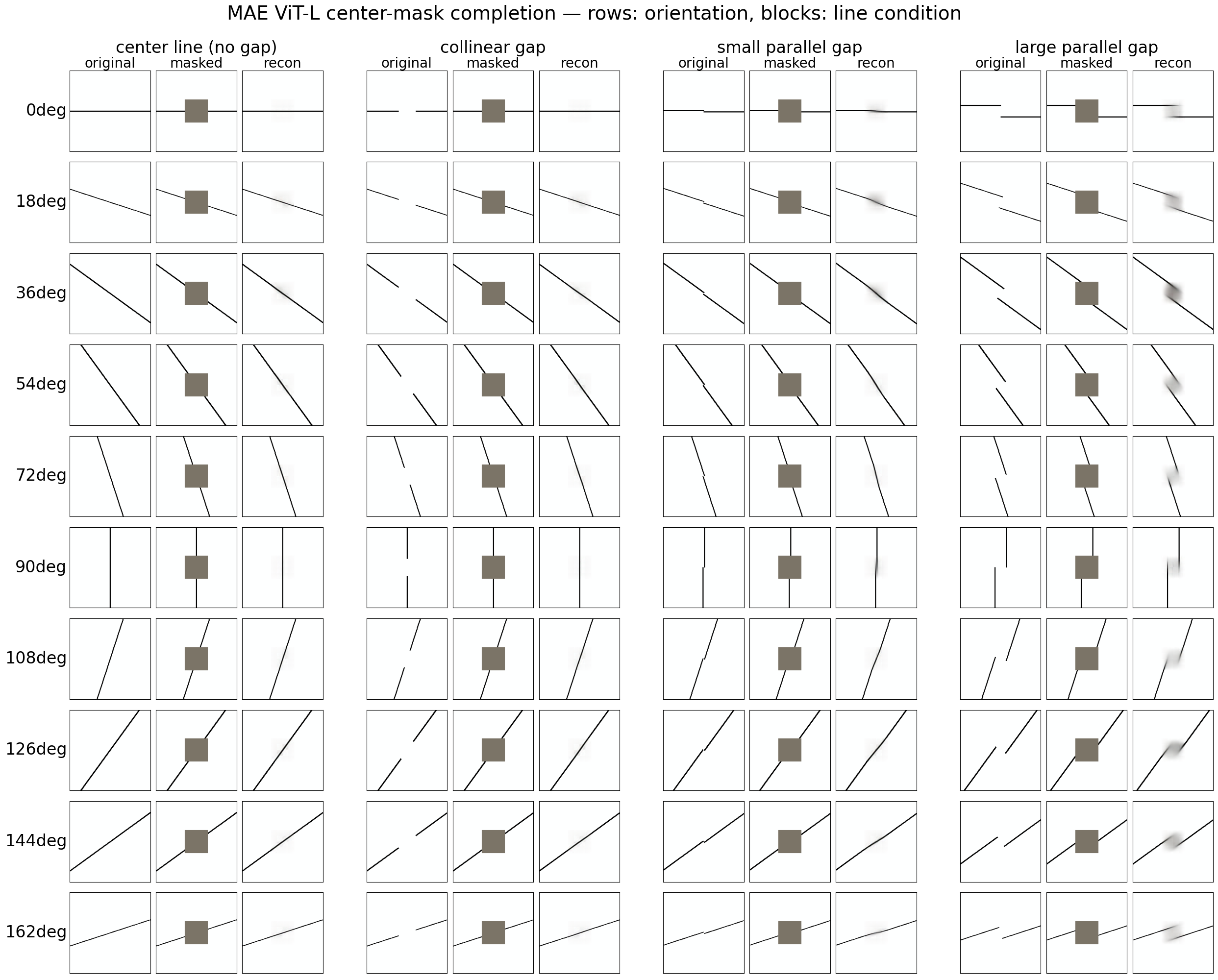}
    \caption{Contour completion across orientations and alignments
    in MAE ViT-L. Rows vary orientation; the four blocks show
    a continuous line, a collinear gap, and small and large
    perpendicular offsets. Each block presents the original
    image, masked input, and reconstruction. Gray squares
    indicate the mask. Collinear inputs generally yield more
    coherent continuations, while larger offsets often produce
    diffuse or displaced marks.}
    \label{fig:mae_line_continuation_comp}
\end{figure}

Figure~\ref{fig:mae_line_continuation_comp} provides a qualitative
comparison of MAE ViT-L reconstructions across stimulus
orientations and four alignment conditions: a continuous line,
a collinear gap, and small and large perpendicular offsets
between parallel segments. A central mask conceals the region
in which continuation is evaluated. These configurations probe
whether reconstructed structure depends on the alignment of
visible segments and whether similar patterns occur across
orientations.

Collinear configurations generally produce more coherent
line-like continuations, whereas larger offsets often yield
diffuse or displaced marks. The examples are consistent with
a preference for good continuation, although reconstruction
quality varies across orientations and configurations.
This comparison illustrates alignment-sensitive completion;
it does not establish exact rotation invariance or a
quantitative threshold for tolerating misalignment.

\subsection{Encoder and decoder contributions to global completion.}
Does adaptation to different image statistics change completion through
the encoder, the decoder, or both?
To examine their contributions, we crossed encoders and complete
decoders from the models continually pretrained for 500 epochs on
natural images or texforms.
We evaluated all four combinations on the same 72 cross-shaped
cue-conflict stimuli, holding the visible input and mask fixed for
each stimulus.
The local reference continues the nearby boundary, whereas the global
reference restores the missing arm to form a symmetric cross.

\begin{figure}[htbp]
    \centering
    \includegraphics[width=0.95\linewidth]{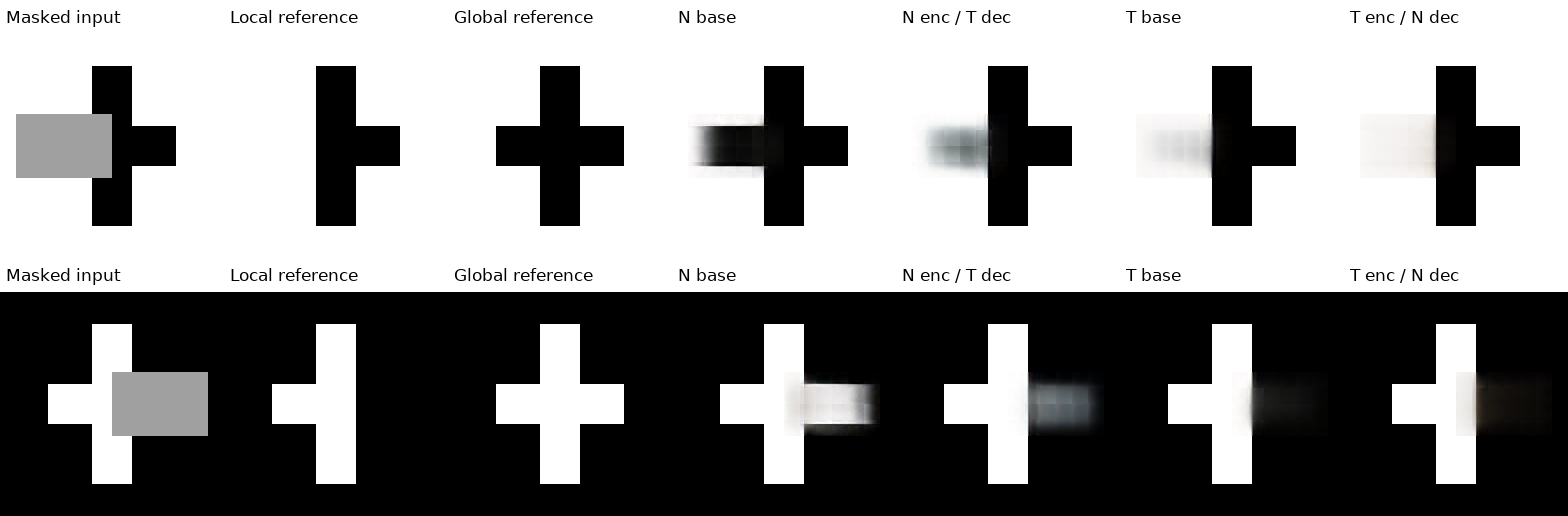}
    \caption{
\textbf{Encoder--decoder swaps alter global completion.}
Columns show the masked input, local and global reference completions,
and reconstructions from four encoder--decoder combinations.
N and T denote models continually pretrained for 500 epochs on natural
images and texforms, respectively; ``base'' denotes the original
encoder--decoder pairing.
Gray indicates the masked region, and the two rows illustrate opposite
contrast polarities.
The natural model reconstructs the missing cross arm, whereas replacing
either component weakens this completion. A faint arm remains when the
natural encoder is paired with the texform decoder.
Table~\ref{tab:encoder_decoder_swaps} reports completion preferences
across all 72 stimulus variants.
}
\label{fig:decoder_encoder_swap_img}
    \label{fig:placeholder}
\end{figure}

Figure~\ref{fig:decoder_encoder_swap_img} illustrates the effects of
exchanging components, and
Table~\ref{tab:encoder_decoder_swaps} quantifies the resulting
completion preferences.
The matched natural-image model favored global completion
(mean $P=+0.831$), whereas the matched texform model favored local
completion ($P=-0.953$).
Keeping the natural encoder but substituting the texform decoder
reversed the mean preference to local completion ($P=-0.589$).
Conversely, pairing the texform encoder with the natural decoder
also produced a strong local preference ($P=-0.974$).
Thus, the natural decoder did not restore global preference when
given texform-encoder representations, and the natural encoder
did not preserve global preference when paired with the texform decoder.

\begin{table}[htbp]
  \centering
  \caption{Cross-stimulus completion preference for encoder--decoder
  combinations. Rows identify the encoder and columns the complete
  decoder, each from a model continually pretrained for 500 epochs.
  Values are mean $\pm$ sample standard deviation across 72 stimulus
  variants. Preference is
  $P=(\mathrm{MSE}_{L}-\mathrm{MSE}_{G})/
  (\mathrm{MSE}_{L}+\mathrm{MSE}_{G})$, evaluated in the region where
  the reference completions disagree. With the P score being 1 denoting the system completes based on the global pattern, whereas P $\rightarrow$ -1 indicating the system prefer to use local continuation to complete. All combinations
  receive identical inputs and masks.}
  \label{tab:encoder_decoder_swaps}
  \begin{tabular}{lcc}
      \hline
      & \multicolumn{2}{c}{Decoder} \\
      \cline{2-3}
      Encoder & Natural & Texform \\
      \hline
      Natural & $+0.831 \pm 0.351$ & $-0.589 \pm 0.368$ \\
      Texform & $-0.974 \pm 0.018$ & $-0.953 \pm 0.054$ \\
      \hline
  \end{tabular}
\end{table}

Among the four tested combinations, a positive mean global preference
was expressed only by the matched natural-image encoder and decoder.
These results show that the completion preference is sensitive to
changes in both components, consistent with contributions from both
the representation of visible structure and its use in reconstruction.
However, component swaps can also disrupt compatibility between
learned representations and their readout.
The results therefore do not by themselves distinguish a loss of
structural information from a change in how that information is used.


\subsection{Grouping-consistent attention in the encoder}
\label{app:encoder_attention}

\begin{figure*}[ht]
    \centering
    \includegraphics[width=0.8\linewidth]{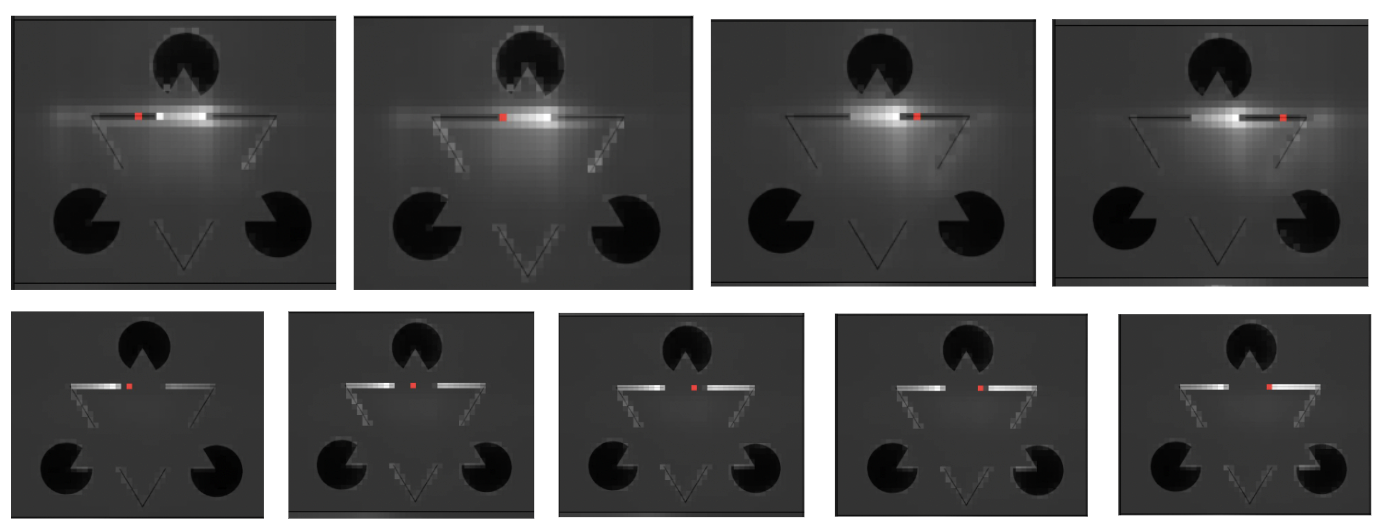}
    \caption{Grouping-consistent attention in the MAE encoder.
    The query token (red) attends to spatially separated regions
    aligned with the implied contour, consistent with good
    continuation. The map illustrates a candidate grouping
    interaction rather than a causal test of completion.}
    \label{fig:attention}
\end{figure*}

Our decoder interventions identify components that contribute
to completion, but grouping-related information may already
be present in the encoder. To explore this possibility,
Figure~\ref{fig:attention} visualizes a selected encoder
attention head for a line-continuation stimulus. The query
token, highlighted in red, attends to spatially separated
regions aligned with the implied contour. This pattern is
consistent with good continuation and suggests that encoder
attention can link elements belonging to a common structure.

This qualitative observation complements the decoder analysis
by illustrating a possible encoder contribution to grouping.
However, attention weights alone do not establish that these
interactions cause completion or distinguish contour grouping
from sensitivity to shared orientation and appearance.

\subsection{Contour completion across convex and concave configurations}
\label{app:convex_concave_completion}

\begin{figure}[ht]
    \centering
    \includegraphics[width=\linewidth]
        {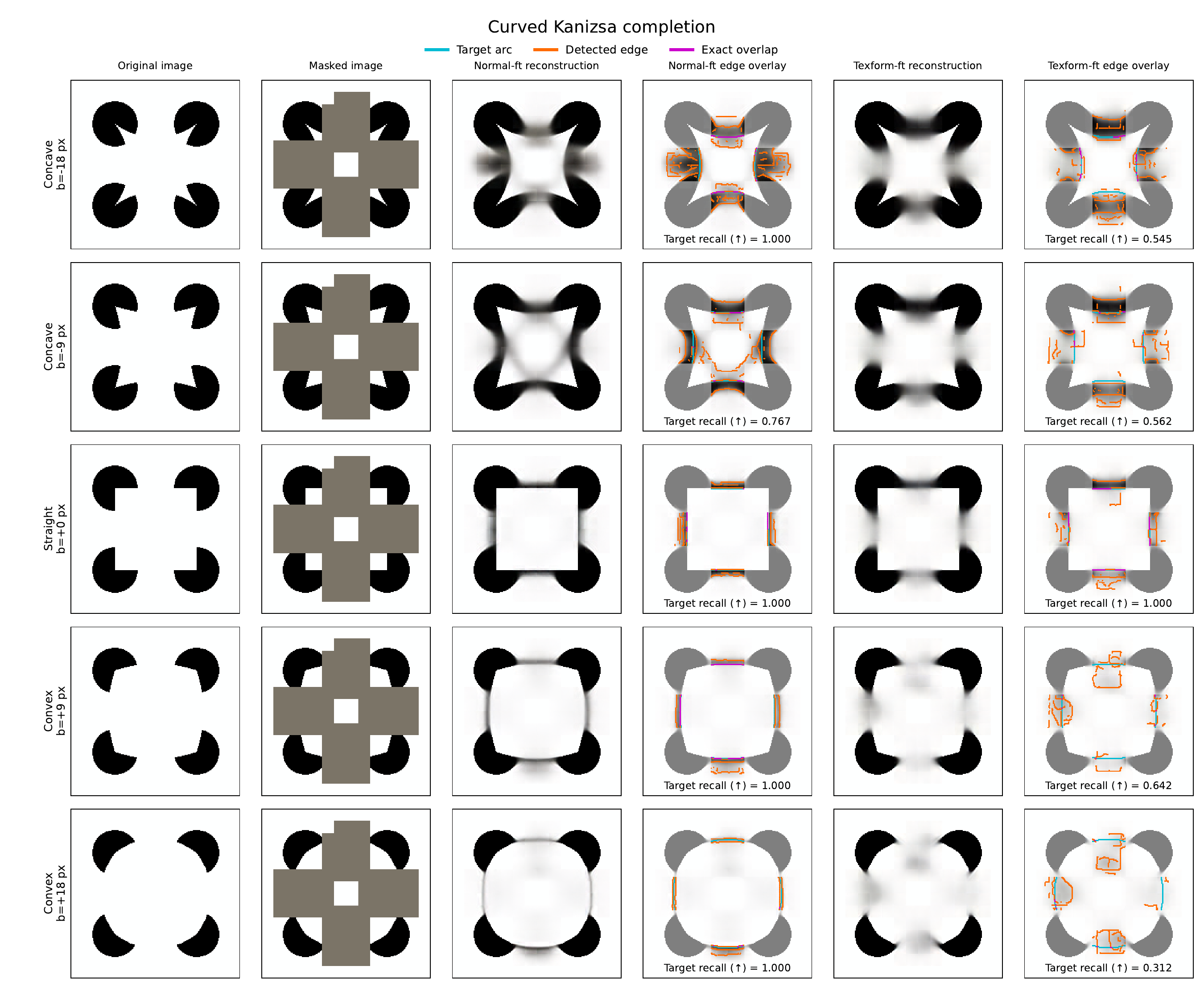}
    \caption{Contour completion across concave, straight, and
    convex configurations at $r=30$ pixels.
    Rows vary boundary bulge from $-18$ to $18$ pixels.
    Columns show the original stimulus, masked input,
    natural-adapted reconstruction and edge overlay, and
    texform-adapted reconstruction and edge overlay.
    Cyan marks target contours, orange detected edges, and
    magenta exact overlap. Recall uses a two-pixel tolerance.
    Natural-image adaptation yields higher recall in all four
    curved examples; both models attain full recall for
    straight boundaries. Additional edges, especially in the
    strongly concave example, illustrate why high recall
    alone does not imply accurate contour reconstruction.}
    \label{fig:convex_concave_ksquare_complete}
\end{figure}

Figure~\ref{fig:convex_concave_ksquare_complete} compares
completion across five boundary geometries at a fixed inducer
radius of $r=30$ pixels. The inter-inducer gap is 48 pixels,
giving a square side length of $s=108$ pixels.
Boundary bulge varies over $b\in\{-18,-9,0,9,18\}$ pixels:
negative values specify concave boundaries relative to the
central region, positive values specify convex boundaries,
and zero specifies straight boundaries. The naturally adapted
and texform-adapted MAEs receive identical visible inputs and
masks for each configuration.

The overlays compare detected reconstruction edges with
target contours defined by the stimulus generator.
Cyan denotes target contours, orange detected edges, and
magenta exact pixel overlap. Target recall measures the
fraction of target pixels with a detected edge within two
pixels; it therefore allows small positional deviations
and does not penalize additional edges.

For the concave configurations, natural-image adaptation
yields target recall of 1.000 and 0.767 at $b=-18$ and
$b=-9$, respectively, compared with 0.545 and 0.562 after
texform adaptation. However, the naturally adapted model
also produces additional edges, particularly in the strongly
concave configuration. These examples indicate greater
coverage of the target contour without implying a clean
or geometrically exact reconstruction.

Both models achieve full target recall for the straight
configuration, although their reconstructions differ in
sharpness and additional detected edges. The convex
configurations show a clearer qualitative separation:
the naturally adapted model reconstructs comparatively
continuous outward-bending boundaries, with recall of
1.000 at both $b=9$ and $b=18$. After texform adaptation,
these boundaries are less distinct and detected edges are
more fragmented or displaced, with recall falling to
0.642 and 0.312, respectively.

These examples show that the effect of texform adaptation
depends on the tested geometry and is not a uniform loss
of boundary reconstruction. They also motivate evaluating
precision, boundary distance, and strength alongside recall.
The configurations test recovery of specified convex and
concave contours; they do not directly measure a preference
between competing convex and concave interpretations or
establish figure--ground assignment.

\subsection{Boundary metrics for curved Kanizsa completion}
\label{app:boundary-metrics}

We quantify reconstructed curved Kanizsa boundaries using boundary F1,
boundary distance, and edge strength. These metrics measure complementary
aspects of the output: proximity and coverage of detected boundaries,
spatial localization error, and intensity variation across the expected
boundary. All geometric distances are measured in pixels of the
$224\times224$ reconstruction.

\paragraph{Image preprocessing and evaluation region.}
We evaluate the reconstructed image formed by inserting predicted pixels
into the masked patches and retaining the visible input elsewhere.
After reversing input normalization, RGB values are multiplied by 255,
clipped to $[0,255]$, and converted to unsigned 8-bit integers. Grayscale
intensity $I$ is the arithmetic mean of the three channels divided by
255, giving values in $[0,1]$.

Scoring is restricted to the interiors of the four inter-inducer gaps.
For inducer radius $r$, let $s=48+2r$ be the side length of the virtual
square and $w=16+\operatorname{round}(s/6)$. The reconstruction mask
selects $16\times16$ patches whose centers lie within $w$ pixels normal
to a side's straight chord and within 36 pixels along that chord from
its midpoint. Within this mask, the evaluation region $\Omega$ retains
pixel centers within $w$ pixels in the normal direction and 22 pixels
in the tangential direction of any side midpoint, and with distance
greater than 6 pixels from the nearest unmasked pixel. The 22-pixel
half-length leaves a 2-pixel margin at each end of the nominal 48-pixel
gap, while the mask-interior restriction excludes reconstruction seams.
For a fixed radius, this region is identical across curvature conditions
and head interventions.

The reference boundary is defined geometrically by the stimulus
generator's four circular arcs. Each arc is sampled at 4,096 points,
rasterized to the nearest pixel center, and the resulting binary curve
is skeletonized and restricted to $\Omega$. Denote this nonempty target
pixel set by $T$. Predicted edges are extracted from the full grayscale
reconstruction using Canny detection with Gaussian smoothing
$\sigma=1$ pixel and fixed low/high thresholds of 0.05/0.10
(non-quantile thresholds). The detected edge map is then restricted to
$\Omega$, giving the predicted pixel set $E$.

\paragraph{Boundary F1: higher is better.}
For a pixel $x$ and a nonempty edge set $A$, define the nearest-boundary
distance as $d(x,A)=\min_{a\in A}\lVert x-a\rVert_2$. Precision and
recall use a tolerance of $\tau=2$ pixels:
\begin{equation}
    P=\frac{1}{|E|}\sum_{e\in E}
        \mathbf{1}\!\left[d(e,T)\leq\tau\right],
    \qquad
    R=\frac{1}{|T|}\sum_{t\in T}
        \mathbf{1}\!\left[d(t,E)\leq\tau\right].
\end{equation}
The boundary F1 score is
\begin{equation}
    F_1=\begin{cases}
        \dfrac{2PR}{P+R}, & P+R>0,\\[3pt]
        0, & P+R=0.
    \end{cases}
\end{equation}
If $E$ is empty, precision, recall, and F1 are all set to zero.
Matching uses nearest-edge proximity independently for each pixel;
it does not impose a one-to-one assignment. F1 lies in $[0,1]$ and
balances target-boundary coverage against detected edges outside the
allowed tolerance.

\paragraph{Boundary distance: lower is better.}
We compute the symmetric mean nearest-boundary distance, with each
individual distance capped at $c=16$ pixels:
\begin{equation}
    D=\frac{1}{2}\left[
        \frac{1}{|E|}\sum_{e\in E}\min\!\left(d(e,T),c\right)
        +\frac{1}{|T|}\sum_{t\in T}\min\!\left(d(t,E),c\right)
    \right].
\end{equation}
This is a capped symmetric Chamfer distance, reported in pixels.
The predicted-to-target term penalizes extraneous or displaced edges,
while the target-to-predicted term penalizes missing coverage. The
cap is applied before averaging. If $E$ is empty, both directional
terms are set to $c$, giving $D=16$. Otherwise $D\in[0,16]$, with zero
indicating coincident rasterized boundary sets.

\paragraph{Edge strength: higher indicates a stronger expected boundary.}
This metric measures image gradients near the known target arcs without
thresholding an edge map. We initially sample each arc at 1,024
approximately equally spaced arc-length positions and estimate unit
normals $\mathbf{n}_i$ from finite differences of the sampled curve.
We retain positions $\mathbf{p}_i$ whose interpolated distance to the
outside of $\Omega$ exceeds 3 pixels, leaving room for a local normal
search. Let
\begin{equation}
    \mathbf{g}(x)=\nabla\bigl(G_{\sigma}*I\bigr)(x),
    \qquad \sigma=1\ \text{pixel},
\end{equation}
where the two gradient components are computed using first derivatives
of a Gaussian. Using bilinear interpolation of these gradient images,
we define
\begin{equation}
    S=\frac{1}{N}\sum_{i=1}^{N}
       \max_{u\in\mathcal{U}}
       \left|\mathbf{g}(\mathbf{p}_i+u\mathbf{n}_i)
       \cdot\mathbf{n}_i\right|,
    \qquad
    \mathcal{U}=\{-2,-1.5,\ldots,1.5,2\}.
\end{equation}
Thus, at each retained arc position, we take the largest absolute
normal derivative within $\pm2$ pixels and then average over positions
pooled across the four sides. The absolute value makes the response
independent of contrast polarity, and the local search tolerates small
spatial offsets. We report $S$ directly in normalized-intensity units
per pixel. Unlike F1 and boundary distance, this measure does not
penalize extra edges elsewhere in the evaluation region; it quantifies
the strength of a nearby intensity transition along the expected normal.

\paragraph{Aggregation and head contributions.}
Each metric is computed separately for each reconstruction and then
averaged with equal weight over the 16 curved stimuli: four radii
$r\in\{24,30,36,42\}$ crossed with four nonzero bulges
$b=\operatorname{round}(s f)$, where
$f\in\{-1/6,-1/12,1/12,1/6\}$. The four straight-boundary controls
are excluded from these curved-stimulus averages. For an individual
head ablation $h$, contributions are the paired effects
$F_{1,0}-F_{1,h}$, $D_h-D_0$, and $S_0-S_h$, averaged over the same
stimuli; subscript 0 denotes the intact model. Positive contributions
indicate deterioration after removing the head. The three metrics
retain their own units and yield separate head rankings; no combined
score is used in the reported metric-specific analyses.

\subsection{Head ablation study for curved Kanizsa square}
\begin{figure}[htbp]
      \centering
      \begin{subfigure}[c]{0.3\linewidth}
          \centering
          \includegraphics[width=\linewidth]{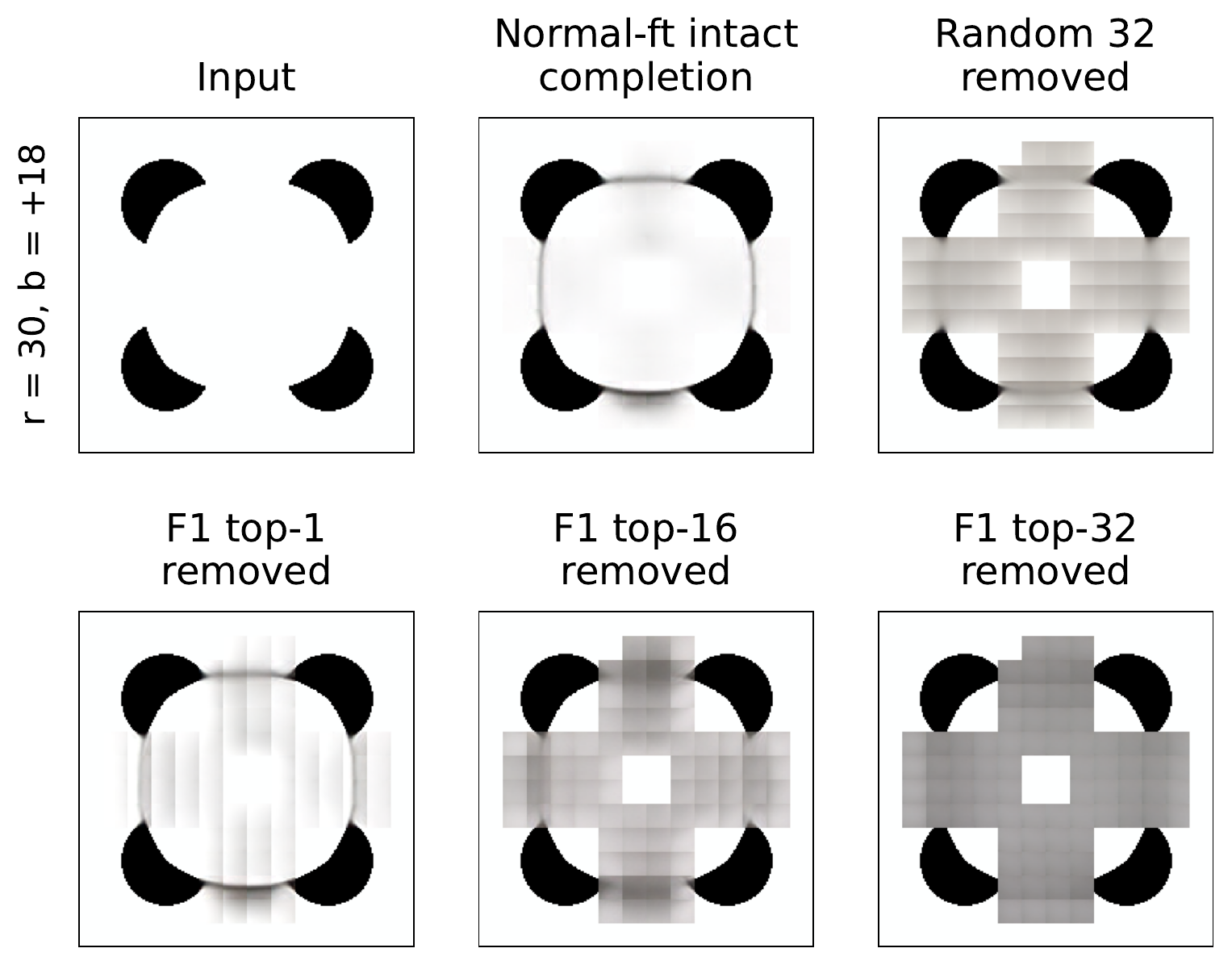}
      \end{subfigure}%
      \hspace{0.015\linewidth}%
      \begin{subfigure}[c]{0.65\linewidth}
          \centering
          \includegraphics[width=\linewidth]{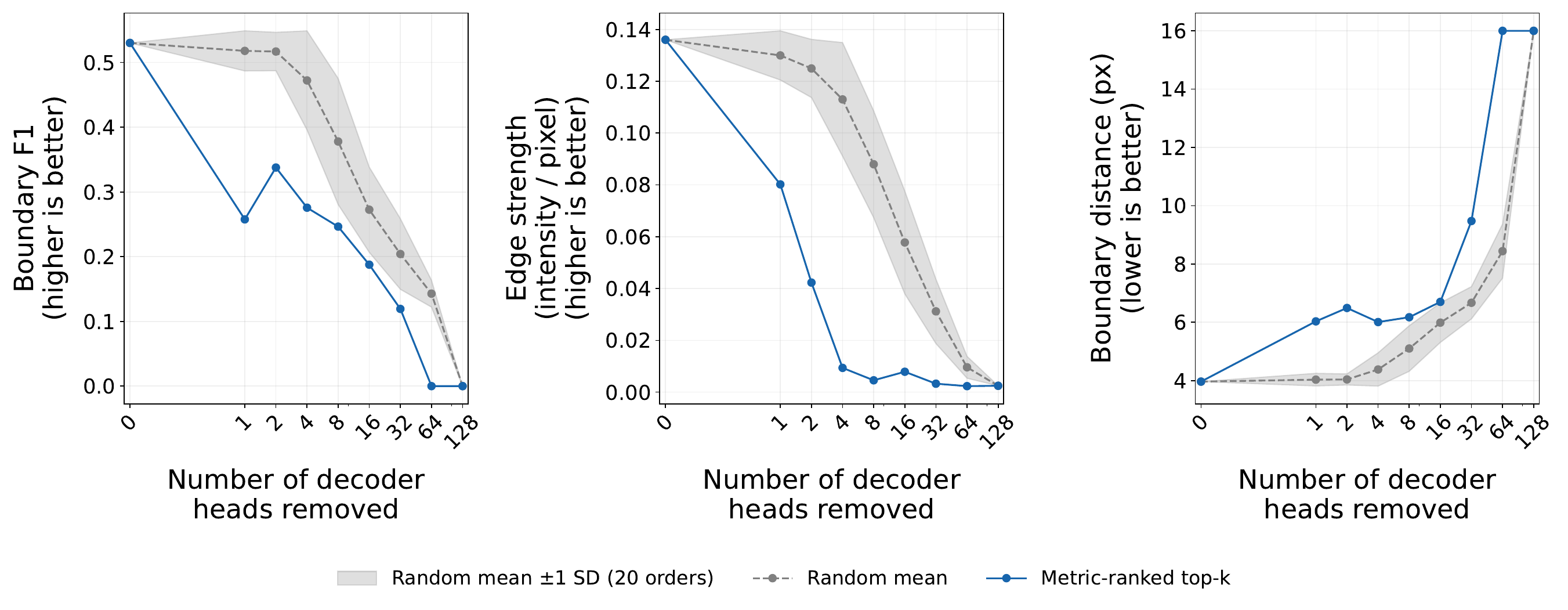}
      \end{subfigure}

      \caption{
      Decoder attention-head ablation in normally trained MAE-B.
      \textbf{Left:} Reconstructions for a curved Kanizsa square with
      $r=30$ and $b=+18$. The first row shows the input, intact-model
      completion, and removal of 32 randomly selected heads. The second
      row shows cumulative removal of the top 1, 16, and 32 heads ranked
      by their individual effects on boundary F1.
      \textbf{Right:} Boundary F1, edge strength, and boundary distance
      as a function of the number of ablated heads, averaged across
      16 curved stimuli. Each panel uses its own metric-specific
      ranking. Blue curves show ranked ablation; gray dashed curves
      and shading indicate the mean and $\pm1$ standard deviation
      across 20 random head orderings. Higher F1 and edge strength,
      and lower boundary distance, indicate better reconstruction.
      }
      \label{fig:ksquare_ablation}
  \end{figure}

To assess the contribution of individual decoder attention heads to curved Kanizsa-square completion, we ablated each of the 128 heads in the normally trained MAE-B by zeroing its attention output before the output projection. Heads were ranked separately by the resulting decrease in boundary F1, decrease in edge strength, or increase in boundary distance, averaged across 16 curved stimuli. We then cumulatively ablated the top-ranked heads and compared the resulting scores with those obtained from 20 random head orderings. As shown in Fig.~\ref{fig:ksquare_ablation}, metric-specific ranked ablation generally impaired completion more strongly than random ablation of the same number of heads. The reconstruction examples illustrate the disruption of the completed boundary as increasing numbers of F1-ranked heads were removed. Because ranking and evaluation used the same stimuli, these results describe head contributions within this stimulus set.

\subsection{Details of the Head ablation study for contour completion}\label{sec:app_more_contour_completion_ablation}
We investigate decoder components that contribute to
contour completion in the normally fine-tuned MAE-B. We rank all
128 decoder attention heads using bidirectional activation
replacement between aligned and disrupted contour conditions
on the original 24 discovery instances. Each head's score
combines the reduction in target reconstruction error when its
aligned activation is transferred to the disrupted condition
and the increase in error under the reverse transfer. This
ranking is fixed before examining held-out reconstructions.

We then ablate the top 1, 8, or 16 ranked heads by setting their
outputs to zero at all decoder token positions, including the
class token, before the attention output projection. We examine
two held-out instances: a vertical contour (instance 29) and a
horizontal contour (instance 60) in Figure~\ref{fig:ranked-decoder-ablation}. The figures show the stimulus
context with the reconstructed target inserted, alongside enlarged
views of the same target patch. Unablated aligned and disrupted
conditions provide reference reconstructions.

In these two examples, removing the top-ranked head produces
relatively little visible change, whereas removing the top 8
or 16 heads weakens the reconstructed contour and increases
target reconstruction error relative to the intact aligned
condition. These examples illustrate the contribution of the
selected decoder heads to contour completion.

Together, activation replacement and zero ablation link
context-dependent reconstruction to specific decoder components.
However, these effects do not establish that the selected heads
are exclusively specialized for contour completion, since the
interventions may also impair general reconstruction.


\subsection{Decoder-head ablations across completion tasks}
\label{app:decoder-head-specialization}

We extend Section~\ref{sec:decoder-head-specialization} by
examining whether the same decoder attention heads support
different completion outcomes. We compare head rankings across
curved Kanizsa completion, local/global cue conflict, and
contour continuation, using each task's existing evaluation
metrics.

\paragraph{Model and ablation procedure.}
We use MAE-B continually pretrained on natural images for
500 epochs and individually ablate each of its 128 decoder
attention heads (eight layers, 16 heads per layer).
Each intervention sets one head's output to zero at all
decoder token positions, including the class token, before
the attention output projection. Every intervention starts
from the intact model, so the effects of different heads
are measured independently rather than cumulatively.

\paragraph{Evaluation stimuli.}
We evaluate 16 curved Kanizsa stimuli, 72 local/global
cue-conflict variants, and 40 aligned contour-continuation
instances (IDs 24--63). The contour experiment uses the
original held-out stimulus set, but the measurements here
are additional single-head ablations. They are distinct
from the preceding cumulative ablation experiment, whose
ranking was obtained by bidirectional activation replacement
on 24 discovery instances. The present rankings are computed
from the single-head ablation effects on these evaluation
sets and are therefore exploratory.

\paragraph{Measuring ablation effects and ranking heads.}
For each head, we measure the paired change relative to
the intact model on each stimulus and average within
each task. Kanizsa completion is evaluated using boundary
F1, boundary distance, and edge strength; cue-conflict
completion uses global preference; and contour continuation
uses target-patch RGB MSE. We express effects as decreases
in F1, edge strength, and global preference, or increases
in boundary distance and MSE. Positive values thus indicate
that removing a head weakens the corresponding measured
outcome. For global preference, this specifically denotes
a shift away from the global reference, rather than
necessarily a reduction in reconstruction accuracy.

Heads are ranked separately for each metric by their mean
ablation effect. We retain the original metric units and
do not combine outcomes into a common score. To assess
whether similar heads influence different outcomes, we
compare rankings using Spearman correlation across all
128 heads and overlap between the top-16 head sets.

\begin{figure*}[ht]
    \centering
    \includegraphics[width=\linewidth]{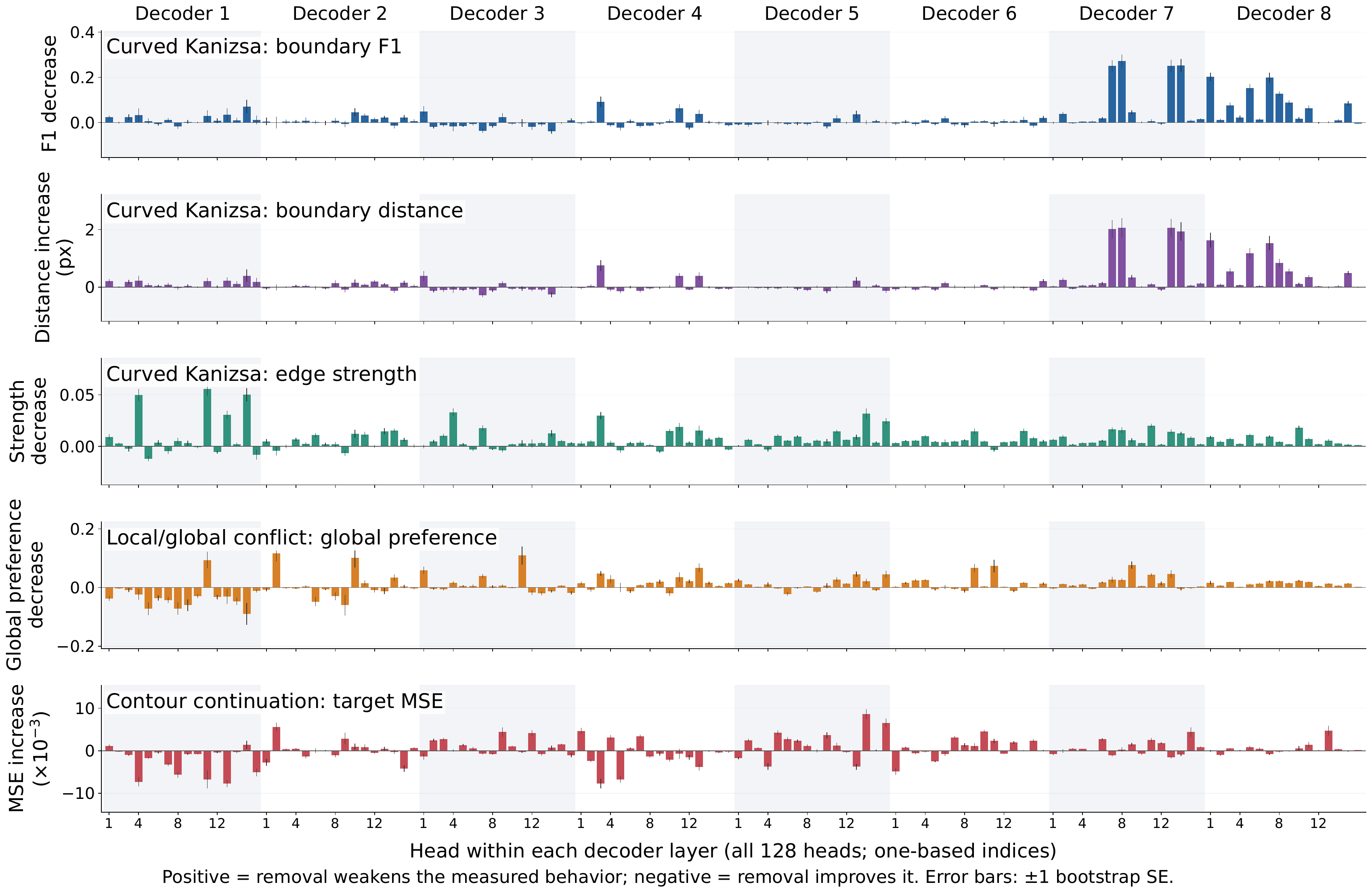}
    \caption{Individual decoder-head contributions in Normal-ft MAE-B
    after 500 epochs. All 128 heads are shown in a common layer/head
    order. From top to bottom, panels report the decrease in Kanizsa
    boundary F1, increase in boundary distance (pixels), decrease in
    edge strength, decrease in global preference on local/global
    cue-conflict stimuli, and increase in contour target-patch RGB MSE
    ($\times 10^{-3}$). Each bar is the mean paired effect of zeroing
    one head at all decoder token positions before the output projection.
    Positive values indicate reduced support for the measured outcome.
    Panels use separate vertical scales. Error bars show $\pm1$ bootstrap
    standard error, computed as the sample standard deviation of means
    across 2,000 within-task stimulus resamples with replacement
    (seed 940). The same sampled indices are used for every head within
    a task and across the three Kanizsa metrics.}
    \label{fig:decoder-head-contributions}
\end{figure*}

\paragraph{Agreement within the Kanizsa task.}
Boundary F1 and boundary distance yield closely aligned
rankings (Spearman $\rho=0.924$), sharing 15 of their top
16 heads. The largest effects for these two metrics are
concentrated in decoder layers 7 and 8. Edge-strength
rankings agree less strongly with either F1
($\rho=0.278$) or distance ($\rho=0.227$). Thus, even
within one stimulus family, heads affecting boundary
contrast and sharpness differ from those most strongly
affecting geometric agreement.

\paragraph{Agreement across tasks.}
Contour-MSE rankings correlate negatively with Kanizsa
F1 ($\rho=-0.215$) and boundary distance ($\rho=-0.258$),
sharing no top-16 heads with either. Global-preference
and contour-MSE rankings also show limited agreement
($\rho=0.102$; two shared top-16 heads).
These results indicate that heads with the largest
effects on one completion outcome are not necessarily
those with the largest effects on another.

\paragraph{Interpretation and limitations.}
The ablations reveal both shared and distinct decoder
contributions across the tested outcomes. Agreement
between the two Kanizsa boundary-accuracy metrics contrasts
with weaker agreement across tasks and with edge strength.
This pattern supports task- and metric-dependent contributions,
but does not identify a dedicated head set for each Gestalt
rule. The tasks differ in stimulus geometry, masks, target
regions, and scoring procedures, and the interventions may
also affect general reconstruction. Moreover, limited
top-16 overlap does not imply that the remaining heads have
no shared effects. These exploratory comparisons are
conditional on the available stimuli and a fixed checkpoint;
they do not establish reproducibility across independent
training runs.
\end{document}